\documentclass[12pt]{JHEP3}
\usepackage{epsfig}
\epsfclipon
\usepackage{multicol}
\usepackage{epsfig,bm}
\usepackage{amssymb,amsmath}
\usepackage{graphicx}

\newcommand{\be}{\begin{equation}}
\newcommand{\ee}{\end{equation}}
\newcommand{\bea}{\begin{eqnarray}}
\newcommand{\beas}{\begin{eqnarray*}}
\newcommand{\eea}{\end{eqnarray}}
\newcommand{\eeas}{\end{eqnarray*}}
\newcommand{\ba}{\begin{array}}
\newcommand{\ea}{\end{array}}

\def\ls{\mathrel{\lower4pt\vbox{\lineskip=0pt\baselineskip=0pt
           \hbox{$<$}\hbox{$\sim$}}}}
\def\gs{\mathrel{\lower4pt\vbox{\lineskip=0pt\baselineskip=0pt
           \hbox{$>$}\hbox{$\sim$}}}}

\def\smiley{\hbox{\large$\bigcirc$\hspace{-.80em}%
\raise.2ex\hbox{$\cdot\cdot$}\kern-.61em    
\lower.2ex\hbox{\scriptsize$\smile$}}\ }

\newcommand{\roughly}[1]{\mathrel{\raise.3ex\hbox{$#1$\kern-0.85em
\lower1ex\hbox{$\sim$}}}}

\def\be{\begin{equation}}
\def\beq\begin{equation}
\def\ee{\end{equation}}
\def\bea{\begin{eqnarray}}
\def\eea{\end{eqnarray}}

\def\beq{\begin{equation}}
\def\eeq{\end{equation}}
\def\beqa{\begin{eqnarray}}
\def\eeqa{\end{eqnarray}}

\newcommand{\bmat}{\left(\begin{array}}
\newcommand{\emat}{\end{array}\right)}

\title{Dressing the inflaton with the Standard Model gauge group}

\author{ A. Mazumdar~$^{1,~2}$\\ $^{1}$~Physics Department, Lancaster
University, LA1 4YB, United Kingdom\\ $^{2}$~ Niels Bohr Institute,
Blegdamsvej-17, Copenhagen-2100, Denmark}

\abstract{In this talk we will discuss how inflation can be embedded
within a minimal extension of the Standard Model where the inflaton
carries the Standard Model charges. There is no need of an ad-hoc
scalar field to be introduced in order to explain the temperature
anisotropy of the cosmic microwave background radiation, all the
ingredients are present within a minimal supersymmetric Standard
Model. The model is robust enough to provide a successful exit from
inflation with all observed matter in the universe. This is a triumph
for an inflationary paradigm which has always begged a simple
question: can we identify the inflaton in a laboratory. We will
briefly discuss how LHC can shed some insight into the inflaton.}

\begin{document}

Inflation has been extremely successful in explaining the temperature
anisotropy of the observed comsic microwave background radiation by
generating almost scale invariant density perturbations~\cite{WMAP3}.
(for a theory of density perturbations see~\cite{Mukhanov-Rev}).  It
has been known for almost 26 years that inflation can be driven by a
dynamical scalar field known as the {\it inflaton}, an order
parameter, which could either be fundamental or
composite. Particularly, if the inflaton rolls very slowly on a
sufficiently flat potential, such that the potential energy density
dominates over the kinetic term, then all the successes of inflation
can be met, i.e. dynamical explanation of the homogeneity and the
isotropy of the universe on very large scales (for a review see
\cite{Lindebook}), Gaussian density perturbations, etc.

However, inspite of the impressive list of achievements, inflation was
never embedded in a fundamental theory which could also be testable in
a laboratory. Inspite of many attempts there has been no single good
candidate for an inflaton which comes naturally out of a well
motivated theory of particle physics (for a review on models of
inflation, see~\cite{Lyth-Riotto}). One always relies on scalar fields
which are {\it absolute gauge singlets} possibly residing in some
hidden sector or secluded sector with a small coupling to the SM gauge
group. By definition an {\it absolute gauge singlet} does not carry
any charge what so-ever be the case. Therefore the masses, couplings
and interactions are not generally tied to any fundamental theory or
any symmetry.  Such gauge singlets are used ubiquitously by model
builders to obtain a desired potential and interactions at a free will
in order to explain the current CMB data. Not only that many notable
papers use ad-hoc couplings to explain phenomena such as preheating
and thermalization without bothering the relevant degrees of
freedom required to create a Universe like ours.

Very recently some of these questions have been addressed in a low
energy field theory setup which explains (for a review see
\cite{anupam}):

\begin{itemize}

\item{ the origin of inflation }

\item{ the fundamental interactions of an inflaton }

\item{ how the inflaton creates Standard Model baryons and cold dark matter ?}

\item{ and  how can we test the inflaton in a laboratory ?}

\end{itemize}

For the first time we are aiming to build a holistic model of
inflation which is truly embedded in a Standard Model (SM) gauge
theory.  The inflaton carries the SM charges and inflation occurs
within an observable sector if the low energy supersymmetry (SUSY) is
found just above the electroweak scale at the
LHC~\cite{AEGM,AEGJM,AKM,AJM,ADM,AFM,AM,ADM2}. We will explain below
why SUSY is necessary to build a successful inflation model. SUSY
provides the scalar fields (partners of the SM fermions and gauge
bosons) and the stability of the flatness of the inflaton potential.

Furthermore, since the inflaton carries the SM charges, it decays {\it
only} into the SM particles and SUSY particles, i.e. quarks, squarks,
leptons, sleptons, etc. Within a minimal supersymmetric Standard Model
(MSSM) we know all the relativistic species and therefore we can trace
back thermal history of the universe accurately above the electroweak
scale. We highlight that all the relevant physical processes are
happening within an observable sector alone. If the lightest
supersymmetric particle (LSP) is stable due to R-parity, we naturally
obtain cold dark matter as a consequence of inflation~\cite{ADM}.


\begin{figure}
\vspace*{-0.0cm}
\begin{center}
\epsfig{figure=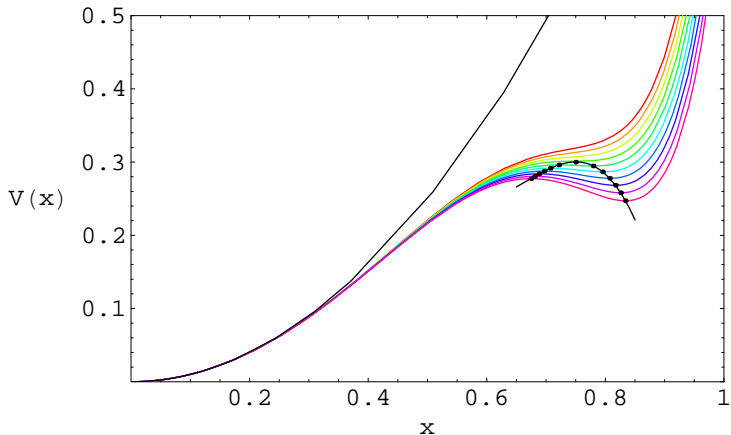,width=.64\textwidth,clip=}
\vspace*{-0.0cm}
\end{center}
\caption{ The colored curves depict the full potential, where
$V(x)\equiv V(\phi)/(0.5~m_{\phi}^2 M_{\rm P}^2(m_{\phi}/M_{\rm
P})^{1/2})$, and $ x\equiv (\lambda_n M_{\rm P}/m_{\phi})^{1/4}
(\phi/M_{\rm P})$. The black curve is the potential arising from the
soft SUSY breaking mass term. The black dots on the colored potentials
illustrate the gradual transition from minimum to the saddle point and
to the maximum.}
\label{fig0-pot}
\end{figure}

Let us now concentrate on the model of inflation which is based on
MSSM together with gravity~\cite{AEGM,AEGJM,ADM}. Therefore
consistency dictates that all non-renormalizable terms allowed by
gauge symmetry and supersymmetry should be included below the cut-off
scale, which we take to be the Planck scale. The superpotential term
which lifts the $F$-flatness is given by (see for a
review~\cite{MSSM-REV}):
\beq \label{supot}
W_{non} = \sum_{n>3}{\lambda_n \over n}{\Phi^n \over M^{n-3}}\,,
\eeq
where $\Phi$ is a {\it gauge invariant} superfield which contains the
flat direction.  Within MSSM all the flat directions are lifted by
non-renormalizable operators with $4\le n\le 9$~\cite{gherghetta96},
where $n$ depends on the flat direction. We expect that quantum
gravity effects yield $M=M_{\rm P}=2.4\times 10^{18}$~GeV and
$\lambda_n\sim {\cal O}(1)$~\cite{dine95}. Note however that
our results will be valid for any values of $\lambda_n$, because
rescaling $\lambda_n$ simply shifts the VEV of the flat direction.
Let us focus on the lowest order superpotential term in
Eq.~(\ref{supot}) which lifts the flat direction. Soft SUSY breaking
induces a mass term for $\phi$ and an $A$-term so that the scalar
potential along the flat direction reads
\beq \label{scpot}
V = {1\over2} m^2_\phi\,\phi^2 + A\cos(n \theta  + \theta_A)
{\lambda_{n}\phi^n \over n\,M^{n-3}_{\rm P}} + \lambda^2_n
{{\phi}^{2(n-1)} \over M^{2(n-3)}_{\rm P}}\,,
\eeq
Here $\phi$ and $\theta$ denote respectively the radial and the
angular coordinates of the complex scalar field
$\Phi=\phi\,\exp[i\theta]$, while $\theta_A$ is the phase of the
$A$-term (thus $A$ is a positive quantity with dimension of mass).
Note that the first and third terms in Eq.~(\ref{scpot}) are positive
definite, while the $A$-term leads to a negative contribution along
the directions whenever $\cos(n \theta + \theta_A) < 0$, see
\ref{fig0-pot}. The importance of the A-term was first highlighted in
a successful MSSM curvaton model~\cite{AEJM}~\footnote{Similar ideas
for the curvaton to carry the SM charges have been entertained
in~\cite{We}.}.

The maximum impact from the $A$-term is obtained when $\cos(n \theta +
\theta_A) = -1$ (which occurs for $n$ values of $\theta$).  For a
choice of parameter, $A^2 = 8 (n-1) m^2_{\phi}$, the potential is flat
near $\phi_0$, where first and second derivative of the potential
vanishes, i.e. $V^{\prime}(\phi_0)=0,~ V^{\prime\prime}(\phi_0)=0$.
As a result, if initially $\phi\sim \phi_0$, a slow roll phase of
inflation is driven by the third derivative of the potential. We will
show that the potential near $\phi_0$ becomes cosmologically flat.  As
a matter of fact in the gravity mediated SUSY breaking case, the
$A$-term and the soft SUSY breaking mass terms are expected to be the
same order of magnitude as the gravitino mass, i.e.  $m_{\phi}\sim A
\sim m_{3/2}\sim {\cal O}(1)~{\rm TeV}$~see~\cite{Nilles}.

If the above condition is not satisfied then for $A^2\geq
8(n-1)m_{\phi}^2$ the potential develops a secondary false minimum
with a charge and color breaking minimum. Although inflation could
occur there~\cite{Guth} but phenomenologically such a situation is not
desired at all. In the other extreme case it is possible to drive
assisted~\cite{assist1} inflation with many flat directions, however,
such a possibility does not arise within MSSM~\cite{Jokinen1}.

The potential near the saddle point, $A^2=8(n-1)m_{\phi}^2$, is very
flat along the {\it real direction} but not along the {\it imaginary
direction}. Along the imaginary direction the curvature is determined
by $m_{\phi}$.  Around $\phi_0$ the field lies in a plateau with a
potential energy
\beq \label{potential}
V(\phi_0) = {(n-2)^2\over2n(n-1)}\,m^2_\phi \phi_0^2
\eeq
with
\beq \label{phi0}
\phi_0 = \left({m_\phi M^{n-3}_{\rm P}\over
\lambda_n\sqrt{2n-2}}\right)^{1/(n-2)}\,.
\eeq
This results in Hubble expansion rate during inflation which is given by
\beq \label{hubble}
H_{\rm inf} = {(n-2) \over \sqrt{6 n (n-1)}} {m_{\phi} \phi_0 \over M_{\rm P}}.
\eeq
When $\phi$ is very close to $\phi_0$, the first derivative is
extremely small. The field is effectively in a de Sitter background,
and we are in self-reproduction (or {\it eternal inflation}) regime
where the two point correlation function for the flat direction
fluctuation grows with time. But eventually classical friction wins
and slow roll begins at $\phi \approx \phi_{\rm
self}$~\cite{AEGM,AEGJM}
\beq \label{self}
(\phi_0-\phi_{\rm self}) \simeq \Big({m_\phi \phi_0^2 \over M_{\rm
P}^3}\Big)^{1/2} \phi_0.
\eeq
The regime of {\it eternal inflation} plays an important role in
addressing the initial condition problem~\cite{AFM}.

The observationally relevant perturbations are generated when $\phi
\approx \phi_{\rm COBE}$. The number of e-foldings between $\phi_{\rm
COBE}$ and $\phi_{\rm end}$, denoted by ${\cal N}_{\rm COBE}$:
\beq \label{cobe}
{\cal N}_{\rm COBE} \simeq {\phi^3_0 \over 2n(n-1)M^2_{\rm P}(\phi_0 -
\phi_{\rm COBE})}.
\eeq
The amplitude of perturbations thus produced is given by~\cite{AEGJM}
\beq \label{ampl}
\delta_{H} \equiv \frac{1}{5\pi}\frac{H^2_{\rm inf}}{\dot\phi} \simeq
\frac{1}{5\pi} \sqrt{\frac{2}{3}n(n-1)}(n-2) ~ \Big({m_\phi M_{\rm P} \over
\phi_0^2}\Big) ~ {\cal N}_{\rm COBE}^2,
\eeq
and the spectral tilt of the power spectrum and its running are found
to be~\cite{AEGM,AEGJM}
\begin{eqnarray}
\label{tilt}
&&n_s = 1 + 2\eta - 6\epsilon \ \simeq \ 1 -
{4\over {\cal N}_{\rm COBE}} \,, \\ \label{running}
&&{d\,n_s\over d\ln k} = - {4\over {\cal N}_{\rm COBE}^2}. \,
\end{eqnarray}

As discussed in~\cite{AEGM,AEGJM}, among nearly 300 flat directions
there are two that can lead to a successful inflation along the lines
discussed above.

One is $udd$ which, up to an overall phase factor, is parameterized by
\beq
\label{example}
u^{\alpha}_i=\frac1{\sqrt{3}}\phi\,,~
d^{\beta}_j=\frac1{\sqrt{3}}\phi\,,~
d^{\gamma}_k=\frac{1}{\sqrt{3}}\phi\,.
\eeq
Here $1 \leq \alpha,\beta,\gamma \leq 3$ are color indices, and $1
\leq i,j,k \leq 3$ denote the quark families. The flatness constraints
require that $\alpha \neq \beta \neq \gamma$ and $j \neq k$.

The other direction is $LLe$~\footnote{When the flat direction
develops a VEV during inflation, it spontaneously breaks $SU(2)\times
U(1)_{y}$, which gives masses to the corresponding gauge bosons.  It
is possible to obtain a seed perturbations for the primordial magnetic
field in this case, see~\cite{EJM}.}, parameterized by (again up to
an overall phase factor)
\beq
L^a_i=\frac1{\sqrt{3}}\left(\begin{array}{l}0\\ \phi\end{array}\right)\,,~
L^b_j=\frac1{\sqrt{3}}\left(\begin{array}{l}\phi\\ 0\end{array}\right)\,,~
e_k=\frac{1}{\sqrt{3}}\phi\,,
\eeq
where $1 \leq a,b \leq 2$ are the weak isospin indices and $1 \leq
i,j,k \leq 3$ denote the lepton families. The flatness constraints
require that $a \neq b$ and $i \neq j \neq k$.  Both these flat
directions are lifted by $n=6$ non-renormalizable
operators~\cite{gherghetta96},
\begin{eqnarray}
W_6\supset\frac{1}{M_{\rm P}^3}(LLe)(LLe)\,,\hspace{1cm}
W_6\supset \frac{1}{M_{\rm P}^3}(udd)(udd)\,.
\end{eqnarray}
The reason for choosing either of these two flat
directions\footnote{Since $LLe$ are $udd$ are independently $D$- and
$F$-flat, inflation could take place along any of them but also, at
least in principle, simultaneously. The dynamics of multiple flat
directions are however quite involved~\cite{Jokinen}.} is twofold: (i)
a non-trivial $A$-term arises, at the lowest order, only at $n=6$; and
(ii) we wish to obtain the correct COBE normalization of the CMB
spectrum.

Those MSSM flat directions which are lifted by operators with
dimension $n=7,9$ are such that the superpotential term contains at
least two monomials, i.e. is of the type
\begin{eqnarray}\label{doesnotcontri}
W \sim \frac{1}{M_{\rm P}^{n-3}}\Psi\Phi^{n-1}\,.
\end{eqnarray}
If $\phi$ represents the flat direction, then its VEV induces a large
effective mass term for $\psi$, through Yukawa couplings, so that
$\langle \psi \rangle =0$. Hence Eq. (\ref{doesnotcontri}) does not
contribute to the $A$-term.

More importantly, as we will see, all other flat directions except
those lifted by $n=6$ fail to yield the right amplitude for the
density perturbations. Indeed, as can be seen in Eq.~(\ref{phi0}), the
value of $\phi_0$, and hence also the energy density, depend on $n$.

According to the arguments presented above, successful MSSM flat direction
inflation has the following model parameters:
\beq
m_{\phi}\sim 1-10~{\rm TeV}\,,~~n=6\,,~~A=\sqrt{40}m_{\phi}\,,
~~\lambda\sim {\cal O}(1)\,.
\label{VALVS}
\eeq
Here we assume that $\lambda$ (we drop the subscript "6") is of order
one, which is the most natural assumption when $M=M_{\rm P}$.

The Hubble expansion rate during inflation and the VEV of the saddle
point are~\footnote{We note that $H_{\rm inf}$ and $\phi_0$ depend
very mildly on $\lambda$ as they are both $\propto \lambda^{-1/4}$.}
\beq \label{values}
H_{\rm inf}\sim 1-10~{\rm GeV}\,,~~~\phi_0 \sim (1-3) \times
10^{14}~{\rm GeV}\,.
\eeq
Note that both the scales are sub-Planckian. The total energy density
stored in the inflaton potential is $V_0 \sim 10^{36}-10^{38}~{\rm
GeV}^4$. The fact that $\phi_0$ is sub-Planckian guarantees that the
inflationary potential is free from the uncertainties about physics at
super-Planckian VEVs. The total number of e-foldings during the slow
roll evolution is large enough to dilute any dangerous relic
away~\cite{AEGJM},
\beq \label{totalefold}
{\cal N}_{\rm tot} \sim 10^3  \,,
\eeq
At such low scales as in MSSM inflation the number of e-foldings,
${\cal N}_{\rm COBE}$, required for the observationally relevant
perturbations, is much less than $60$~\cite{BURGESS-multi}.  If
the inflaton decays immediately after the end of inflation, we obtain
${\cal N}_{\rm COBE} \sim 50$. Despite the low scale, the flat
direction can generate adequate density perturbations as required to
explain the COBE normalization. This is due to the extreme flatness of
the potential (recall that $V'=0$), which causes the velocity of the
rolling flat direction to be extremely small. From Eq.~(\ref{ampl}) we
find an amplitude of
\beq
\label{amp}
\delta_{H} \simeq 1.91 \times 10^{-5}\,.
\eeq

There is a constraint on the mass of the flat direction from the
amplitude of the CMB anisotropy:
\begin{equation}
\label{mbound} m_{\phi} \simeq (100 ~ {\rm GeV}) \times \lambda^{-1}
\, \left( \frac{{\cal N}_{\rm COBE}}{50} \right)^{-4}\,.
\end{equation}
We get a lower limit on the mass parameter when $\lambda\leq 1$.
For smaller values of $\lambda\ll 1$, the mass of the flat
direction must be larger.  Note that the above bound on the inflaton
mass arises at high scales, i.e. $\phi=\phi_0$. However, through
renormalization group flow, it is connected to the low scale mass, as
will be discussed in Sect. 4.

The spectral tilt of the power spectrum is not negligible because,
although the first slow roll parameter is $\epsilon\sim1/{\cal N}_{\rm
COBE}^4\ll 1$, the other slow roll parameter is given by $\eta =
-2/{\cal N}_{\rm COBE}$ and thus, see
Eq.~(\ref{tilt})\footnote{Obtaining $n_s > 0.92$ (or $n_s < 0.92$,
which is however outside the $2 \sigma$ allowed region) requires
deviation from the saddle point condition, $A^2=8(n-1)m_{\phi}^2$, see
the discussion below. For a more detailed discussion on the spectral
tilt, see also Refs.~\cite{LYTH1},\cite{AM}.}
\begin{eqnarray}
\label{spect}
&&n_s
\sim 0.92\,,\\
&&{d\,n_s\over d\ln k}
\sim - 0.002\,,
\end{eqnarray}
where we have taken ${\cal N}_{\rm COBE} \sim 50$ (which is the
maximum value allowed for the scale of inflation in our model). In the
absence of tensor modes, this agrees with the current WMAP 3-years'
data within $2\sigma$~\cite{WMAP3}. Note that MSSM inflation does not
produce any large stochastic gravitational wave background during
inflation. Gravity waves depend on the Hubble expansion rate, and in
our case the energy density stored in MSSM inflation is very small.
Inflation can still happen for small deviations from the saddle point
condition, $A^2=8(n-1)m_{\phi}^2$. Notable point is that the spectral
tilt can match the current observations, i.e. within $0.92 \leq
n_s\leq 1.0$. The plot above summarizes the results.

\begin{figure}[t]
\vspace{2cm}
\includegraphics[width=6.5cm]{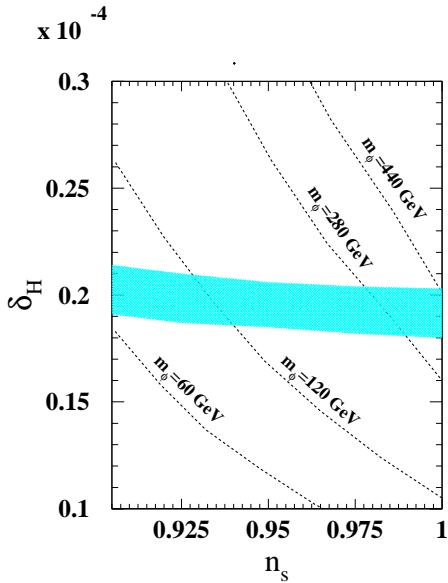}
\caption{$\delta_H$ is plotted as a function of $\Delta^2$ for
different values of $m_{\phi}$. We used $\lambda$ =1. The blue band
denotes the experimentally allowed values of $\delta_H$. Note that the
parameter space allows the spectral tilt to be within $0.92 \leq
n_s\leq 1.0$.}
\label{nsdel}
\end{figure}

The blue band above shows the experimentally allowed region. We find
that smaller values of $m_{\phi}$ are preferred for smaller values of
$n_s$. We also find that the allowed range of $m_{\phi}$ is
$75-440$~GeV for the experimental ranges of $n_s$ and $\delta_H$.  We
assume $\lambda\sim 1$ for these two figures. If $\lambda$ is less
than ${\cal O}(1)$, e.g., $\lambda \sim 0.1$ or so (which can occur in
$SO(10)$ model), it will lead to an increase in $m_{\phi}$.

Since the MSSM inflaton candidates are represented by {\it gauge
invariant} combinations which are not singlets. The inflaton
parameters receive corrections from gauge interactions which, unlike
in models with a gauge singlet inflaton, can be computed in a
straightforward way.  Quantum corrections result in a logarithmic
running of the soft supersymmetry breaking parameters $m_\phi$ and
$A$. The conclusion is robust, although the soft terms and the value
of the saddle point are all affected by radiative corrections, they do
not remove the saddle point nor shift it to unreasonable values. The
existence of a saddle point is thus insensitive to radiative
corrections.

One final comment is in order before closing this Section. Unlike
$m_{\phi}$, there is no prospect of measuring the $A$ term, because it
is related to the non-renormalizable interactions which are suppressed
by $M_{\rm P}$. However, a knowledge of supersymmetry breaking sector
and its communication with the observable sector may help to link the
non-renormalizable $A$-term under consideration to the renormalizable
ones. For instance in the Polonyi model the non-renormalizable
$A$-term and the trilinear $A$-term can be related to each other:
$A_6={3 - \sqrt{3} \over 6 - \sqrt{3}} A_3$.

SUGRA corrections often destroy the slow roll predictions of
inflationary potentials; this is the notorious SUGRA-$\eta$
problem~\cite{ETA}. In general, the effective potential depends on the
K\"ahler potential $K$ as $ V\sim
\left(e^{K(\varphi^{\ast},\varphi)/M_{\rm P}^2} V(\phi)\right) $ so
that there is a generic SUGRA contribution to the flat direction
potential of the type
\begin{equation}
\label{mflat}
V(\phi)=H^2M_{\rm P}^2 f\left(\frac{\phi}{M_{\rm P}}\right)\,,
\end{equation}
where $f$ is some function (typically a polynomial).  Such a
contribution usually gives rise to a Hubble induced correction to the
mass of the flat direction with an unknown coefficient, which depends
on the nature of the K\"ahler potential~\footnote{If the K\"ahler
potential has a shift symmetry, then at tree level there is no Hubble
induced correction. However, at one-loop level relatively small Hubble
induced corrections can be induced~\cite{GMO,ADM1}.}.

Let us compare the non-gravitational contribution, Eq.~(\ref{scpot}),
to that of Hubble induced contribution, Eq.~(\ref{mflat}). Writing
$f\sim \left( \phi/M_{\rm P}\right)^p$ where $p\ge 1$ is some power,
we see that non-gravitational part dominates whenever
\beq
H_{\rm inf}^2M_{\rm P}^2\left(\frac{\phi}{M_{\rm P}}\right)^p \ll
m_{\phi}^2\phi_0^2\,,
\eeq
so that the SUGRA corrections are negligible as long as $\phi_0 \ll
M_{\rm P}$, as is the case here (note that $H_{\rm inf} M_{\rm P} \sim
m_{\phi} \phi_0$).  The absence of SUGRA corrections is a generic
property of this model. Note also that although non-trivial K\"ahler
potentials give rise to non-canonical kinetic terms of squarks and
sleptons, it is a trivial exercise to show that at sufficiently low
scales, $H_{\rm inf}\ll m_{\phi}$, and small VEVs, they can be rotated
to a canonical form without affecting the potential~\footnote{The same
reason, i.e. $H_{\rm inf}\ll m_{\phi}$ also precludes any large
Trans-Planckian correction. Any such correction would generically go
as $(H_{\rm inf}/M_{\ast})^2\ll 1$, where $M_{\ast}$ is the scale at which
one would expect Trans-Planckian effects to kick in.}.

After the end of inflation, the flat direction starts rolling towards
its global minimum. At this stage the dominant term in the scalar
potential will be: $m_\phi \phi^2/2$. Since the frequency of
oscillations is $\omega \sim m_{\phi} \sim 10^3 H_{\rm inf}$, the flat
direction oscillates a large number of times within the first Hubble
time after the end of inflation. Hence the effect of expansion is
negligible.

We recall that the curvature of the potential along the angular
direction is much larger than $H^2_{\rm inf}$. Therefore, the flat
direction has settled at one of the minima along the angular direction
during inflation from which it cannot be displaced by quantum
fluctuations. This implies that no torque will be exerted, and hence
the flat direction motion will be one dimensional, i.e. along the
radial direction.

Flat direction oscillations excite those MSSM degrees of freedom which
are coupled to it.  The inflaton, either ${LLe}$ or ${udd}$ flat
direction, is a linear combination of slepton or squark
fields. Therefore inflaton has gauge couplings to the gauge/gaugino
fields and Yukawa couplings to the Higgs/Higgsino fields. As we will
see particles with a larger couplings are produced more copiously
during inflaton oscillations. Therefore we focus on the production of
gauge fields and gauginos. Keep in mind that the VEV of the MSSM flat
direction breaks the gauge symmetry spontaneously, for instance ${\bf
udd}$ breaks $SU(3)_C \times U(1)_Y$ while ${\bf LLe}$ breaks
$SU(2)_{W}\times U(1)_{Y}$, therefore, induces a supersymmetry
conserving mass $\sim g \langle \phi(t) \rangle$ to the gauge/gaugino
fields in a similar way as the Higgs mechanism, where $g$ is a gauge
coupling. When the flat direction goes to its minimum, $\langle
\phi(t)\rangle = 0$, the gauge symmetry is restored. In this respect
the origin is a point of enhanced symmetry~\cite{AVERDI2}.

There can be various phases of particle creation in this model, here
we briefly summarize the most dominant one.  Let us elucidate the
physics, by considering the case when ${LLe}$ flat direction is the
inflaton ~\footnote{ Reheating happens quickly due to a flat direction
motion which is {\it strictly} one dimensional in our case. Our case
is really exceptional, usually, the flat direction motion is
restricted to a plane, which precludes preheating all together, for
instance see~\cite{Rouz-Camp,Longeviety}.}.

An efficient bout of particle creation occurs when the inflaton
crosses the origin, which happens twice in every oscillation. The
reason is that fields which are coupled to the inflaton are massless
near the point of enhanced symmetry. Mainly electroweak gauge fields
and gauginos are then created as they have the largest coupling to the
flat direction. The production takes place in a short interval,
$\Delta t \sim \left(g m_{\phi} \phi_0 \right)^{-1/2}$, where
$\phi_0\sim 10^{14}$~GeV is the initial amplitude of the inflaton
oscillation, during which quanta with a physical momentum $k \ls
\left(g m_{\phi} \phi_0 \right)^{1/2}$ are produced. The number
density of gauge/gaugino degrees of freedom is given
by~\cite{PREHEAT}, see also~\cite{Cormier}
\beq \label{chiden}
n_{g} \approx {\left(g m_{\phi} \phi_0
\right)^{3/2} \over 8 \pi^3}\,.
\eeq
As the inflaton VEV is rolling back to its maximum value $\phi_0$, the
mass of the produced quanta $g \langle \phi(t) \rangle$ increases. The
gauge and gaugino fields can (perturbatively) decay to the fields
which are not coupled to the inflaton, for instance to (s)quarks. Note
that (s)quarks are not coupled to the flat direction, hence they
remain massless throughout the oscillations. The total decay rate of
the gauge/gaugino fields is then given by $\Gamma = C \left(g^2/48\pi
\right) g\phi $, where $C \sim {\cal O}(10)$ is a numerical factor
counting for the multiplicity of final states.

The decay of the gauge/gauginos become efficient when~\cite{AEGJM}
\beq \label{ddecay}
\langle \phi \rangle \simeq \left({48 \pi m_{\phi} \phi_0 \over
C g^3}\right)^{1/2}\,.
\eeq
Here we have used $\langle \phi(t) \rangle \approx \phi_0 m_{\phi} t$,
which is valid when $m_{\phi} t \ll 1$, and $\Gamma \simeq t^{-1}$,
where $t$ represents the time that has elapsed from the moment that
the inflaton crossed the origin. Note that the decay is very quick
compared with the frequency of inflaton oscillations, i.e. $\Gamma \gg
m_{\phi}$. It produces relativistic (s)quarks with an
energy~\cite{AEGJM}:
\beq \label{energy}
E =\frac{1}{2}g\phi(t)
\simeq \left({48 \pi m_{\phi} \phi_0 \over C g}\right)^{1/2}\,.
\eeq
The ratio of energy density in relativistic particles thus produced
$\rho_{rel}$with respect to the total energy density $\rho_0$ follows
from Eqs.~(\ref{chiden},\ref{energy}):
\beq
\label{ratio}
{\rho_{rel} \over \rho_0} \sim 10^{-2} g\,,
\eeq
where we have used $C \sim {\cal O}(10)$.  This implies that a
fraction $\sim {\cal O}(10^{-2})$ of the inflaton energy density is
transferred into relativistic (s)quarks every time that the inflaton
passes through the origin. This is so-called instant preheating
mechanism~\cite{INSTANT}~\footnote{In a favorable condition the flat
direction VEV coupled very weakly to the flat direction inflaton could
also enhance the perturbative decay rate of the
inflaton~\cite{ABM}.}. It is quite an efficient mechanism in our model
as it can convert almost all of the energy density in the inflaton into
radiation within a Hubble time (note that $H^{-1}_{\rm inf} \sim 10^3
m^{-1}_{\phi}$).

A full thermal equilibrium is reached when ${\it a)~kinetic }$ and
${\it b)~chemical~equilibrium }$ are established. The maximum
(hypothetical) temperature attained by the plasma would be given by:
\beq
\label{tmax}
T_{max} \sim V^{1/4} \sim \left(m_{\phi}\phi_0\right)^{1/2}
\geq 10^{9}~{\rm GeV}\,.
\eeq
This temperature may be too high and could lead to thermal
overproduction of gravitinos~\cite{Ellis}. However the
dominant source of gravitino production in a thermal bath is
scattering which include an on-shell gluon or gluino leg. However
there exists a natural solution to this problem and we showed that the
final reheat temperature is actually well below Eq.~(\ref{tmax}),
i.e. $T_{R}\ll T_{max}$.

One comment is in order before closing this subsection. The gravitinos
can also be created non-perturbatively during inflaton oscillations,
both of the helicity $\pm 3/2$~\cite{MAROTO} and helicity $\pm 1/2$
states~\cite{REST}. In models of high scale inflation (i.e. $H_{\rm
inf} \gg m_{3/2}$) helicity $\pm 1/2$ states can be produced very
efficiently (and much more copiously than helicity $\pm 3/2$ states).
At the time of production these states mainly consist of the inflatino
(inflaton's superpartner).  However these fermions also decay in the
form of inflatino, which is coupled to matter with a strength which is equal
to that of the inflaton. Therefore, they inevitably decay at a similar
rate as that of inflaton, and hence pose no threat to primordial
nucleosynthesis~\cite{MAR}.

In the present case $m_{\phi} \sim m_{3/2} \gg H_{\rm inf}$. Therefore
low energy supersymmetry breaking is dominant during inflation, and
hence helicity $\pm 1/2$ states of the gravitino are not related to
the inflatino (which is a linear combination of leptons or quarks)at
any moment of time. As a result helicity $\pm 1/2$ and $\pm 3/2$
states are excited equally, and their abundances are suppressed due to
kinematical phase factor.  Moreover there will be no dangerous
gravitino production from perturbative decay of the inflaton
quanta~\cite{AVERDI1,AVERDI3,SURFACE}. The reason is that the
inflaton is not a {\it gauge singlet} and has gauge strength couplings
to other MSSM fields. This makes the $inflaton \rightarrow inflatino
~+~ gravitino$ ~decay mode totally irrelevant.


Let us briefly discuss the cold dark matter issue. Note that our model
of inflation is embedded within MSSM, it is a bonus that the cold dark
matter candidate comes out quite naturally once we assume the
R-parity. In a simple toy model like mSUGRA there are only four
parameters and one sign. These are $m_0$ (the universal scalar soft
breaking mass at the GUT scale $M_{\rm G}$); $m_{1/2}$ (the universal
gaugino soft breaking mass at $M_{\rm G}$); $A_0$ (the universal
trilinear soft breaking mass at $M_{\rm G}$)~\footnote{The
relationship between the two $A$ terms, the trilinear, $A_0$ and the
non-renormalizable $A$ term in Eq.(\ref{scpot}) can be related to each
other, however, that depends on the SUSY breaking sector. For a
Polonyi model, they are given by:
$A=(3-\sqrt{3})/(6-\sqrt{3})A_0$~\cite{AEGJM}.}; $\tan\beta = \langle
H_2 \rangle \langle H_1 \rangle$ at the electroweak scale (where $H_2$
gives rise to $u$ quark masses and $H_1$ to $d$ quark and lepton
masses); and the sign of $\mu$, the Higgs mixing parameter in the
superpotential ($W_{\mu} = \mu H_1 H_2$).  Unification of gauge
couplings within supersymmetry suggests that $M_{\rm G} \simeq 2
\times 10^{16}$ GeV. The model parameters are already significantly
constrained by different experimental results.  In subsequent plots we
show that there exists an interesting overlap between the constraints
from inflation and the CDM abundance~\cite{ADM}.

\begin{figure}[t]
\vspace{1cm} 
\includegraphics[width=8.0cm]{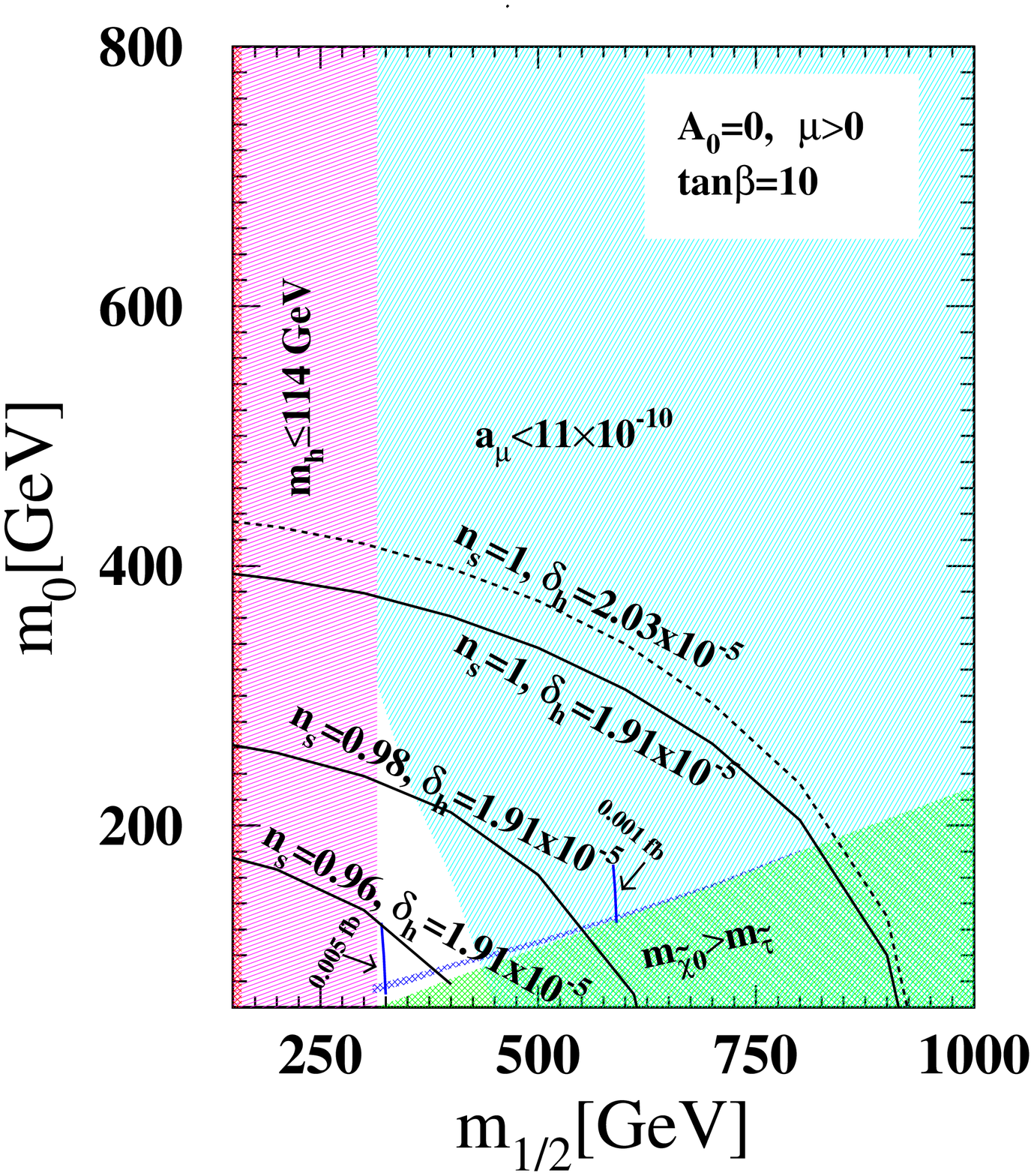}
\includegraphics[width=8.0cm]{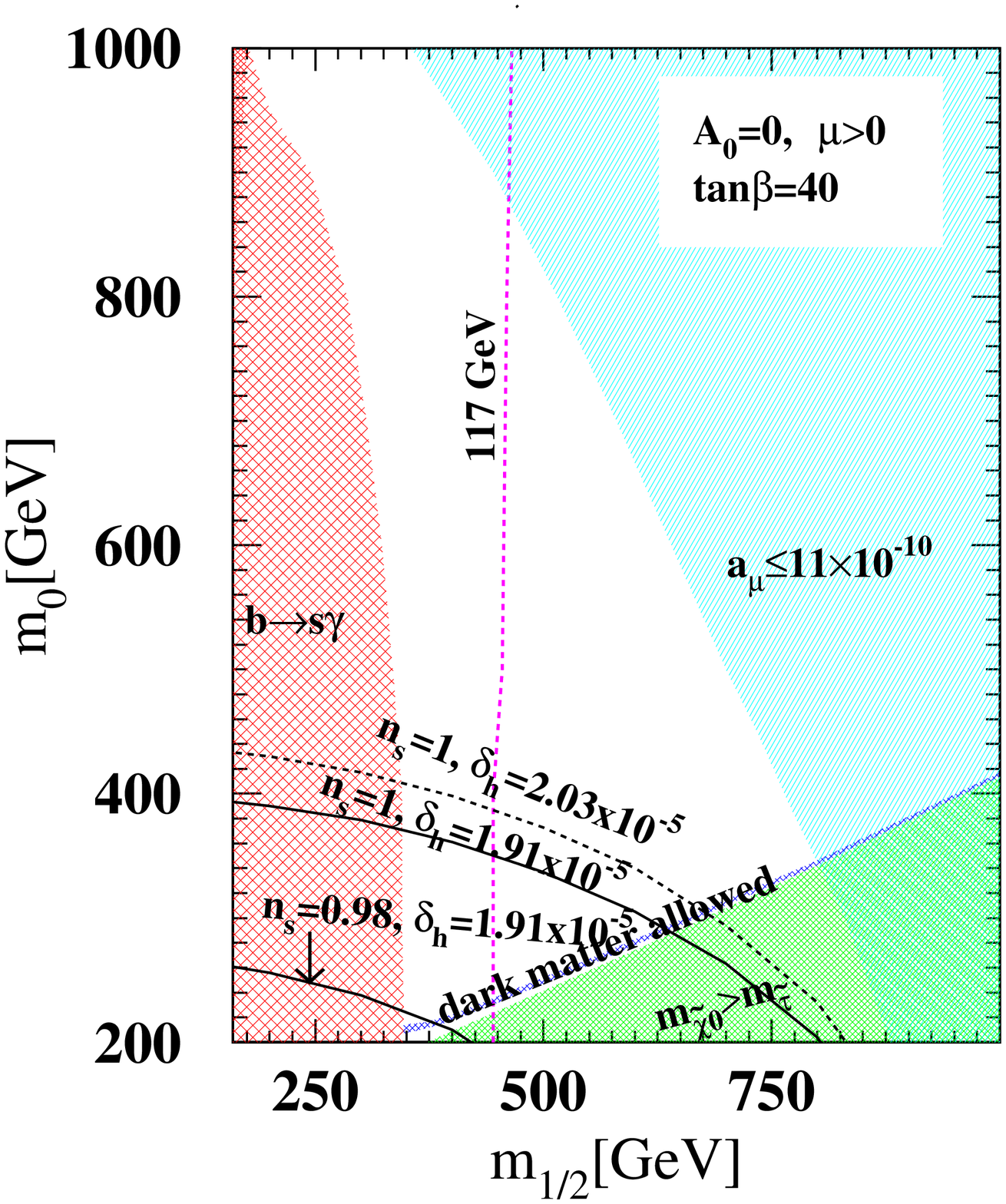}
\caption{The contours for different values of $n_s$ and $\delta_H$ are
shown in the $m_0-m_{1/2}$ plane for $\tan\beta=10$ and $\tan\beta
=40$. We used $\lambda=1$ for the contours. We show the dark matter
allowed region {narrow blue corridor}, (g-2)$_\mu$ region (light blue)
for $a_{\mu}\leq 11\times10^{-8}$, Higgs mass $\leq 114$ GeV (pink
region) and LEPII bounds on SUSY masses (red). We also show the the
dark matter detection rate by vertical blue lines.
} \label{10flat}
\end{figure}


\begin{figure}[t]
\vspace{1cm} 
\includegraphics[width=8.0cm]{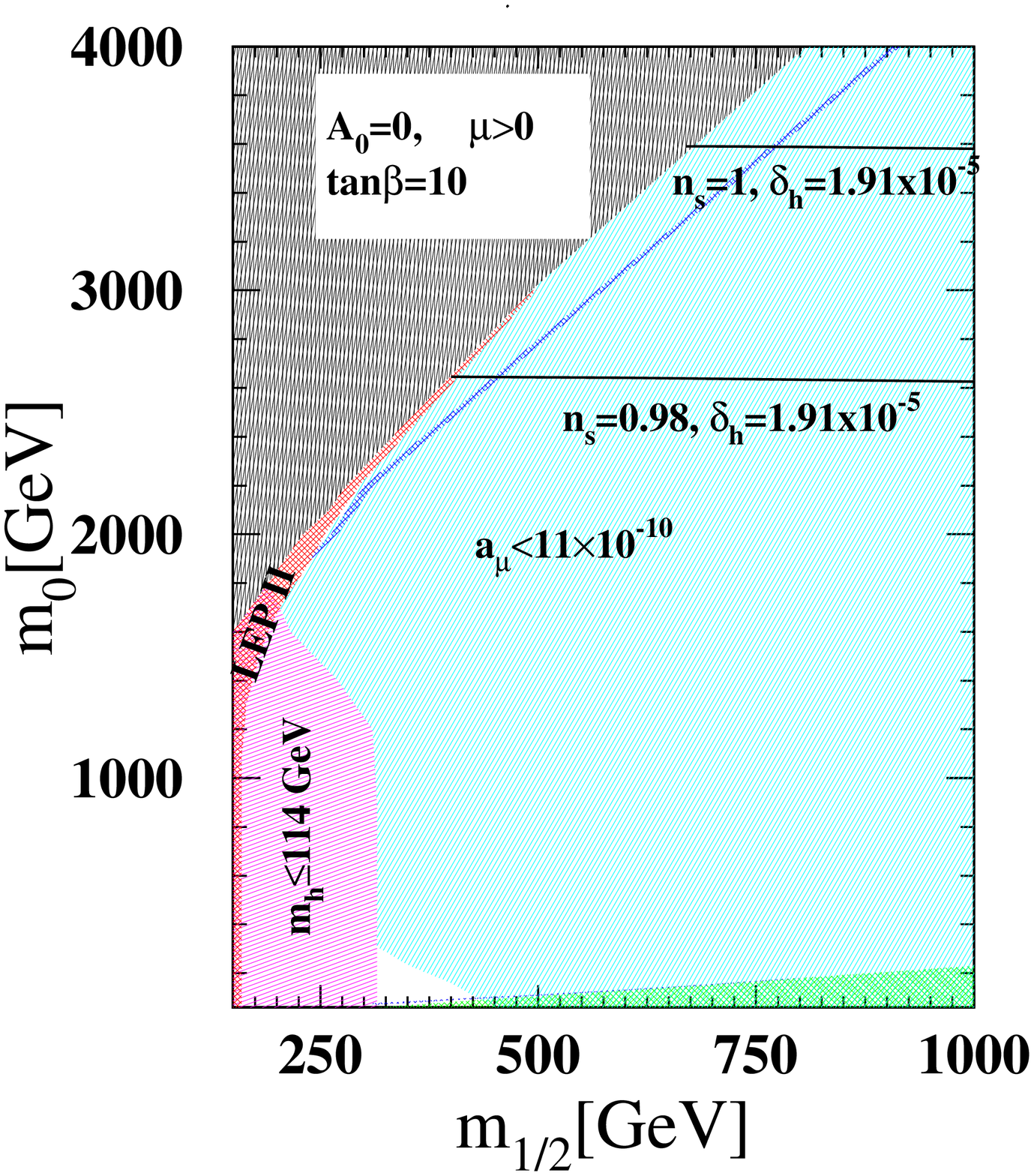}
\includegraphics[width=8.0cm]{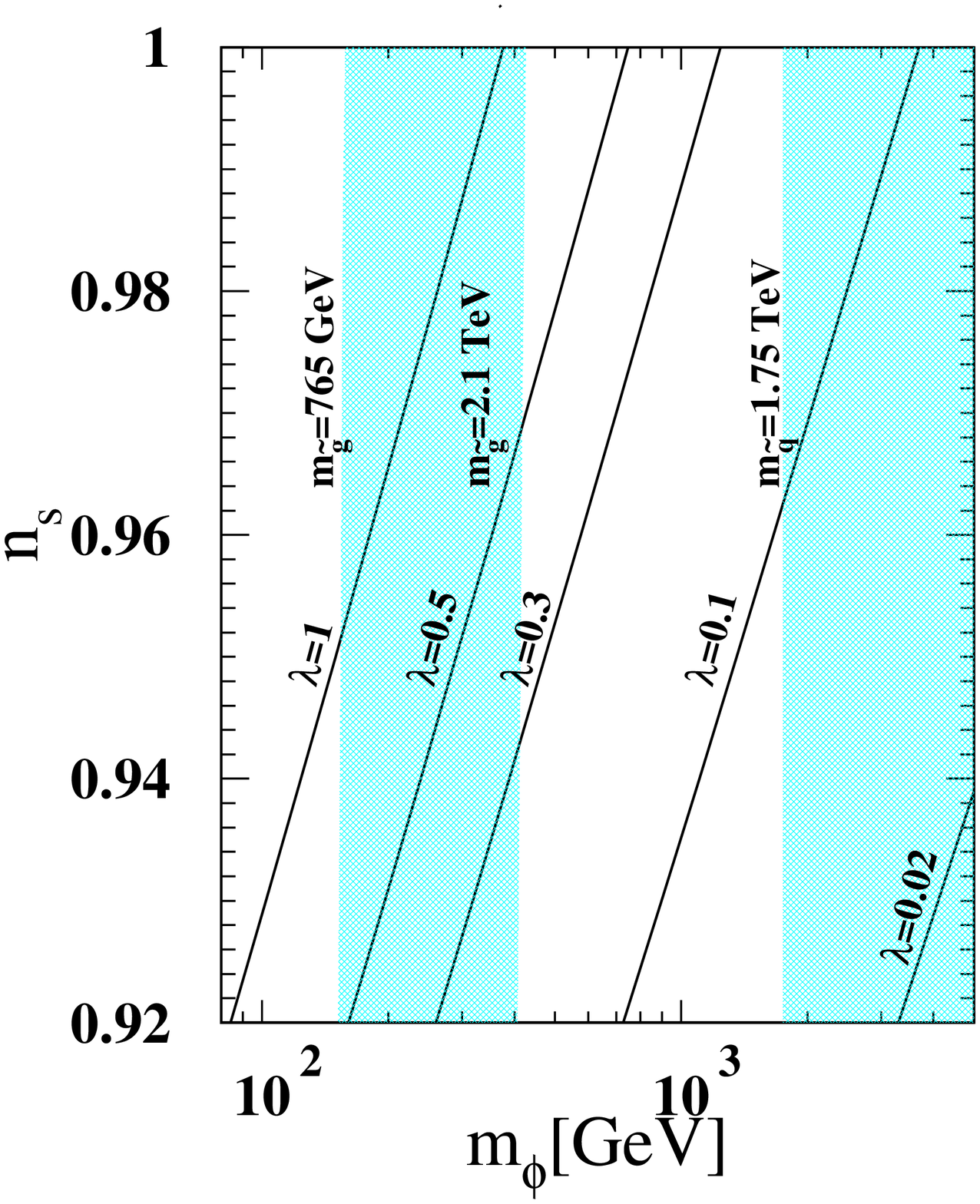}
\caption{The contours for different values of $n_s$ and $\delta_H$
are shown in the $m_0-m_{1/2}$ plane for $\tan\beta=10$. We used
$\lambda=0.1$ for the contours. We show the dark matter allowed
region {narrow blue corridor}, g-2 region (light blue) for
$a_{\mu}\leq 11\times10^{-8}$, Higgs mass $\leq 114$ GeV (pink
region) and LEPII bounds on SUSY masses (red). The black region is
not allowed by radiative electroweak symmetry breaking. We use
$m_t=172.7$~GeV for this graph. In the right hand side plot we show the
contours of $\lambda$  for $\delta_H=1.91\times 10^{-5}$ in
the $n_s$-$m_{\phi}$ plane. The blue band on the left is due to the
stau-neutralino coannihilation region for $\tan\beta=10$ and the
blue band on the right (which continues beyond the plotting range)
denotes the focus point region.} \label{10flatfocus}
\end{figure}



We show that the mSUGRA parameter space in figures~\ref{10flat},
for $\tan\beta=10$ and $40$ with the $udd$ flat direction
using $\lambda=1$~\footnote{We have a similar figure for the flat
direction $LLe$ which we do not show in this paper. All the figures
are for $udd$ flat direction as an inflaton.}. In the figures, we show
contours correspond to $n_s=1$ for the maximum value of
$\delta_H=2.03\times 10^{-5}$ (at $2\sigma$ level) and
$n_s=1.0,~0.98,~0.96$ for $\delta_H=1.91\times 10^{-5}$. The
constraints on the parameter space arising from the inflation
appearing to be consistent with the constraints arising from the dark
matter content of the universe and other experimental results.  We
find that $\tan\beta$ needs to be smaller to allow for smaller values
of $n_s<1$. It is also interesting to note that the allowed region of
$m_{\phi}$, as required by the inflation data for $\lambda=1$ lies in
the stau-neutralino coannihilation region which requires smaller
values of the SUSY particle masses. The SUSY particles in this
parameter space are, therefore, within the reach of the LHC very
quickly. The detection of the region at the LHC has been considered in
refs~\cite{dka}. From the figures, one can also find that as
$\tan\beta$ increases, the inflation data along with the dark matter,
rare decay and Higgs mass constraint allow smaller ranges of
$m_{1/2}$. For example, the allowed ranges of gluino masses are 765
GeV-2.1 TeV and 900 GeV-1.7 TeV for $\tan\beta=10$ and 40
respectively~\cite{ADM}.

So far we have chosen $\lambda=1$. Now if $\lambda$ is small e.g.,
$\lambda\ls 10^{-1}$, we find that the allowed values of $m_{\phi}$
to be large. In this case the dark matter allowed region requires
the lightest neutralino to have larger Higgsino component in the
mSUGRA model. As we will see shortly, this small value of $\lambda$
is accommodated in $SO(10)$ type model. In figure~\ref{10flatfocus},
we show $n_s=1,\, 0.98$ contours for $\delta_H=1.91\times 10^{-5}$
in the mSUGRA parameter space for $\tan\beta=10$. In this figure, we
find that $n_s$ can not smaller than 0.97, but if we lower $\lambda$
which will demand larger $m_{\phi}$ and therefore $n_s$ can be
lowered down to 0.92.

In the second panel of figure~\ref{10flatfocus}, we show the contours
of $\lambda$ for different values of $m_{\phi}$ which are allowed by
$n_s$ and $\delta_H=1.91\times 10^{-3}$. The blue bands show the dark
matter allowed regions for $\tan\beta=10$. The band on the left is due
to the stau-neutralino coannihilation region allowed by other
constraints and the allowed values of $\lambda$ are 0.3-1.  The first
two generation squarks masses are 690 GeV and 1.9 TeV for the minimum
and maximum values of $m_{\phi}$ allowed by the dark matter and other
constraints. The gluino masses for these are $765$~GeV and $2.1$~TeV
respectively.  The band is slightly curved due to the shifting of
$\phi_0$ as a function $\lambda$. (We solve for SUSY parameters from
the inflaton mass at $\phi_0$). The band on the right which continues
beyond the plotting range of the figure~\ref{lamcon} is due to the
Higgsino dominated dark matter. We find that $\lambda$ is mostly $\leq
0.1$ in this region and $m_{\phi}>1.9$ TeV. In this case the squark
masses are much larger than the gluino mass since $m_0$ is much larger
than $m_{1/2}$.


There have been other scintillating developments in embedding
inflation, particularly, realizing saddle point inflation in particle
physics. The flatness of the inflaton potential can be accounted for a
weakness of the Dirac Yukawa couplings for the observed neutrino mass
spectrum~\cite{AKM,ADM2}. In Ref.~\cite{ADM2}, we provided an example
where part of the inflaton flat direction acts as a thermal dark
matter candidate while part of it decays into the SM baryons. The
detection of dark matter and the neutrino properties in neutrino-less
double beta decay experiments would shed important lights on the
inflaton origin.

To summarize, in near future it will be possible to unveil the origin
of the inflaton in a terrestrial laboratory such as the LHC and the
neutrino-less double beta decay experiments. Recognizing the inflaton
dressed with a SM gauge group will be the most cherished success of an
inflationary paradigm.


The research of A.M. is partly supported by the European Union through
Marie Curie Research and Training Network ``UNIVERSENET''
(MRTN-CT-2006-035863).


\end{document}